\documentclass[aps,prd,twocolumn,amsmath,amssymb,amsfonts,nofootinbib,superscriptaddress,altaffilletter]{revtex4-1}

\usepackage{graphicx}
\usepackage{hyperref}
\hypersetup{
  bookmarksopen=true
}
\usepackage{ulem}
\usepackage{amssymb}
\usepackage{amsmath}
\usepackage{color}
\usepackage{supertabular}
\usepackage{dcolumn}
\usepackage[mathscr]{eucal}
\usepackage{mathrsfs}

\def\sci#1#2{#1\times10^{#2}}

\def\RAJ{\textrm{RA}_{\textrm J2000}}
\def\DECJ{\textrm{DEC}_{\textrm J2000}}

\begin{document}

\title{Search for continuous gravitational waves from small-ellipticity sources at low frequencies}

\author{Vladimir Dergachev}
\email{vladimir.dergachev@aei.mpg.de}
\affiliation{Max Planck Institute for Gravitational Physics (Albert Einstein Institute), Callinstrasse 38, 30167 Hannover, Germany}
\affiliation{Leibniz Universit\"at Hannover, D-30167 Hannover, Germany}

\author{Maria Alessandra Papa}
\email{maria.alessandra.papa@aei.mpg.de}
\affiliation{Max Planck Institute for Gravitational Physics (Albert Einstein Institute), Callinstrasse 38, 30167 Hannover, Germany}
\affiliation{Leibniz Universit\"at Hannover, D-30167 Hannover, Germany}
\affiliation{University of Wisconsin Milwaukee, 3135 N Maryland Ave, Milwaukee, WI 53211, USA}

\begin{abstract}
We present the results of an all-sky search for continuous gravitational wave signals with frequencies in the 20-500\,Hz range from neutron stars with ellipticity of $\approx 10^{-8}$. This frequency region is particularly hard to probe because of the quadratic dependence of signal strength on frequency. The search employs the Falcon analysis pipeline \cite{Dergachev:2019wqa} on LIGO O2 public data. Compared to previous Falcon analyses the coherence length has been quadrupled, with a corresponding increase in sensitivity. This enables us to search for small ellipticity neutron stars in this low frequency region up to 44\,pc away. The frequency derivative range is up to $\sci{3}{-13}$\,Hz/s easily accommodating sources with ellipticities of $10^{-7}$ at a distance of a few hundred parsecs. New outliers are found, many of which we are unable to  associate with any instrumental cause.
\end{abstract}

\maketitle

\section{Introduction}

Detectable continuous gravitational waves are expected from fast rotating neutron stars if they present some sort of non-axially symmetric deformation. The deformation in this context is usefully described by the {\it {ellipticity}} of the object, defined as $I_{zz}/ (I_{xx}-I_{yy})$, where $I$ is the moment of inertia tensor of the star and $\hat{z}$ is along the star's rotation axis \cite{Jaranowski:1998qm}. 

In our previous papers \cite{O2_falcon, O2_falcon2} we searched for gravitational wave emission in the 500-2000\,Hz range, targeting objects with ellipticity of $10^{-8}$. 
We found a number of outliers, some corresponding to known instrumental artifacts, some in data with pristine spectrum (i.e., no visible artifacts). 

Ellipticities of $10^{-8}$ are interesting: they are well within the range of what the neutron star crusts can sustain \cite{Gittins:2020cvx} and observational indications exist that millisecond pulsars might have a {\it{minimum}} ellipticity of $\approx 10^{-9}$ \cite{ellipticity}. 

There have been two recent papers \cite{population1, population2} with models of populations of potentially detectable neutron stars. 

In \cite{population1} the authors conclude that the most likely detection will happen in the frequency range covered by this search. Their conclusions are somewhat pessimistic due in part to a number of simplifying assumptions. In particular they assume an exponential model of frequency evolution, rather than a power law that one expects when spin down is dominated by gravitational or electromagnetic emission. We want to highlight that it is reasonable to make simplifications when starting work on such a complicated question, and with further study the predictions will get refined.

In \cite{population2} the authors favor higher frequencies and higher spindown rates than we have searched. This is due to the focus of paper \cite{population2} on ``probing'' neutron stars at a given ellipticity, without taking into account the frequency evolution of neutron stars. This approach is also reasonable, because finding a high-ellipticity source would be very interesting and thus it is important to know which parameter space to search. 

However, finding a high-ellipticity and high frequency source is difficult. A source with a given ellipticity will lose energy due to the emission of gravitational waves and electromagnetic braking. The gravitational wave amplitude grows with the square of the rotation frequency. An isolated high frequency ($\sim 1000$\,Hz) and high ellipticity source ($\sim 10^{-6}$) would have a fairly rapid frequency drift ($\sim -10^{-8}$\,Hz/s) just due to emission of gravitational waves alone. At such spindown rate its frequency will decrease by $100$\,Hz in the span of only 300\,years. Thus we would need to be very lucky to be observing a high-ellipticity isolated neutron star at the very beginning of its spin down evolution.

Sources with smaller ellipticities and smaller spin rates, on the other hand, evolve more slowly. In this paper we explore these sources: with ellipticities of $10^{-8}$ and $10^{-7}$ and spin frequencies below $250$\,Hz, corresponding to the $\ell=m=2$ gravitational wave mode emission at twice the rotation frequency. In addition, the distribution of known radio pulsars has many more sources in this range.

The spindown of the population of known pulsars has a considerably wider spread in our target frequency range,  compared to our previous low-ellipticity searches \cite{O2_falcon, O2_falcon2}. While there are known pulsars with frequency drift on the order of $10^{-10}$\,Hz/s, most are well within $\sci{3}{-13}$\,Hz/s which is the frequency drift tolerance used by this search.

\begin{figure}[htbp]
\includegraphics[width=3.3in]{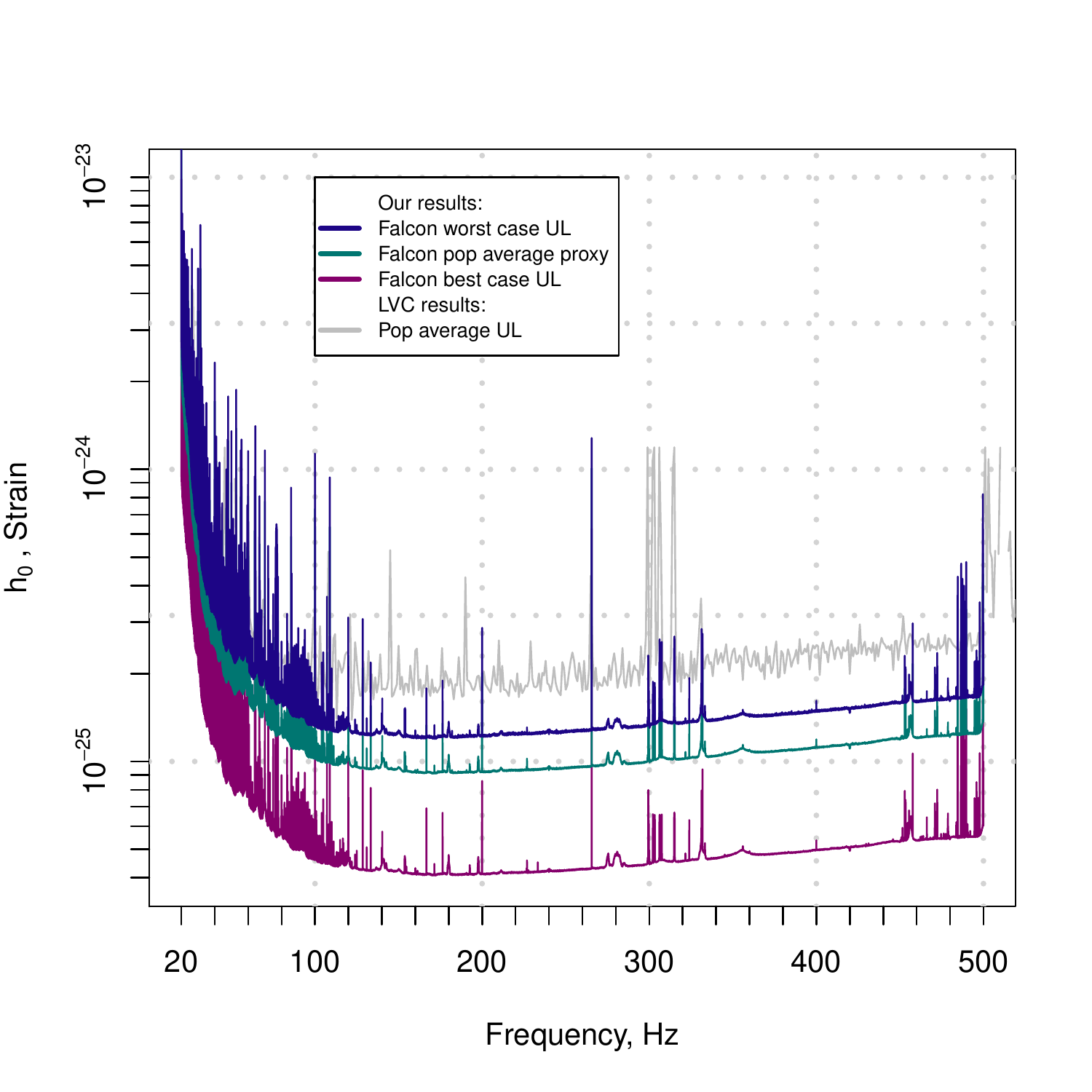}
\caption[Upper limits]{
\label{fig:amplitudeULs}
Gravitational wave intrinsic amplitude $h_0$ upper limits at 95\% confidence as a function of signal frequency. The upper limits are a measure of the sensitivity of the search. We compare with the latest LIGO/Virgo all-sky results in this frequency range \cite{lvc_O2_allsky}, which are a factor $\gtrsim$ 2 less constraining than ours, albeit able to detect sources with much larger distortions. 
}
\end{figure}

Similarly to \cite{deep}, we make a tradeoff between breadth and depth of search. Compared to \cite{lvc_O2_allsky,EHO2}, this search concentrates on low ellipticity/low spindown objects and achieves a factor of 2 and 30\% better sensitivity (Fig. \ref{fig:amplitudeULs}), respectively, than \cite{lvc_O2_allsky} and \cite{EHO2}. Compared to \cite{deep}, that explored 1 Hz, our frequency range covers the entire $20-500$\,Hz space. The improvements to the Falcon pipeline \cite{Dergachev:2019wqa, loosely_coherent, loosely_coherent2, loosely_coherent3} in computing speed allow 2\,day coherence length for the first analysis stage, a factor of 4 longer than used in our previous searches \cite{O2_falcon, O2_falcon2}.

\section{The search and the parameter space}

We use LIGO O2 open data \cite{losc,o2_data, o2_data2}. Our search looks for signals that are phase coherent over a certain time-span, called the coherence-length, and that evolve slowly over longer timescales. The search detects with highest efficiency IT2 signals. In \cite{O2_falcon2} we introduced the notion of the ``IT2 model signal'' for continuous gravitational wave signals 
because in the literature this signal model is used as a reference to measure search performance, and for upper limits. These are signals 
with constant amplitude and a frequency evolution that can be described by a Taylor expansion of order 2 around nominal frequency and frequency derivative values, see for example Section II of \cite{Jaranowski:1998qm}. The ``I'' indicates the source being isolated and ``T2'' indicates an intrinsic frequency evolution given by a Taylor polynomial of degree 2.

Our target frequency range of 20-500\,Hz is particularly difficult to analyze because of a multitude of detector artifacts, including combs below 100\,Hz.  Furthermore,  for the same physical distortion of the neutron star, the putative signals are weaker at these frequencies, compared to higher frequencies.

\begin{figure}[htbp]
\includegraphics[width=3.3in]{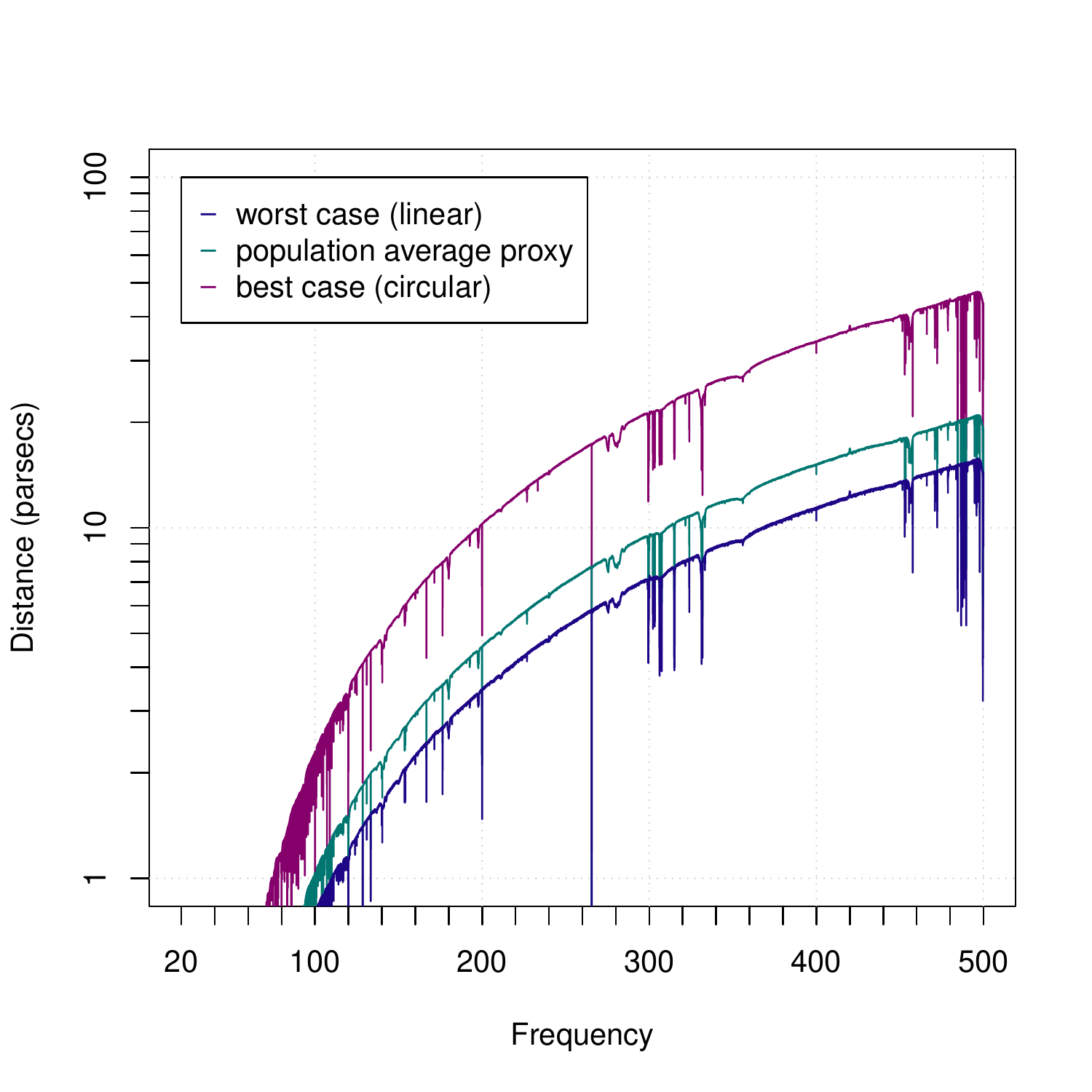}
\caption[Spindown range]{
\label{fig:distance}
Reach of the search for stars with ellipticity of $10^{-8}$. The search is also sensitive to sources with ellipticities of $10^{-7}$ with a distance from Earth that is 10 times higher. The X axis is the gravitational wave frequency, which is twice the pulsar rotation frequency for emission due to an equatorial ellipticity. R-modes and other emission mechanisms give rise to emission at different frequencies. The top curve (purple) shows the reach for a population of circularly polarized sources; The middle curve (cyan) holds for a population of sources with random orientations; The bottom curve (blue) holds for linearly polarized sources. }
\end{figure}

In order to overcome this, we increase the sensitivity of the Falcon pipeline by using long coherence lengths at all stages of the search, as shown in Table~\ref{tab:pipeline_parameters}. Regions of parameter space associated with high SNR results are searched again with progressively longer coherence length.  

With these choices we would be able to see sources with ellipticity of $10^{-8}$ up to 44\,pc away (Figure \ref{fig:distance}). The frequency derivative search range ($\pm \sci{3}{-13}$\,Hz/s) has been chosen to accommodate sources with ellipticities of $10^{-7}$. These sources could be seen up to a distance of $440$ pc.

\begin{table}[htbp]
\begin{center}
\begin{tabular}{rD{.}{.}{2}D{.}{.}{3}}\hline
Stage & \multicolumn{1}{c}{Coherence length (days)} & \multicolumn{1}{c}{Minimum SNR}\\
\hline
\hline
1  & 2 & 6 \\
2  & 3 & 9 \\
3  & 4 & 11 \\
4  & 6 & 12 \\
5  & 9 & 16 \\
6  & 16 & 16 \\
7  & 16 & 16 \\
\hline
\end{tabular}
\end{center}
\caption{Parameters for each stage of the search. Stage 7 refines outlier parameters by using denser sampling of parameter space, and then subjects them to an additional consistency check by comparing outlier parameters from analyses of individual interferometer data.}
\label{tab:pipeline_parameters}
\end{table}

\begin{figure}[htbp]
\includegraphics[width=3.3in]{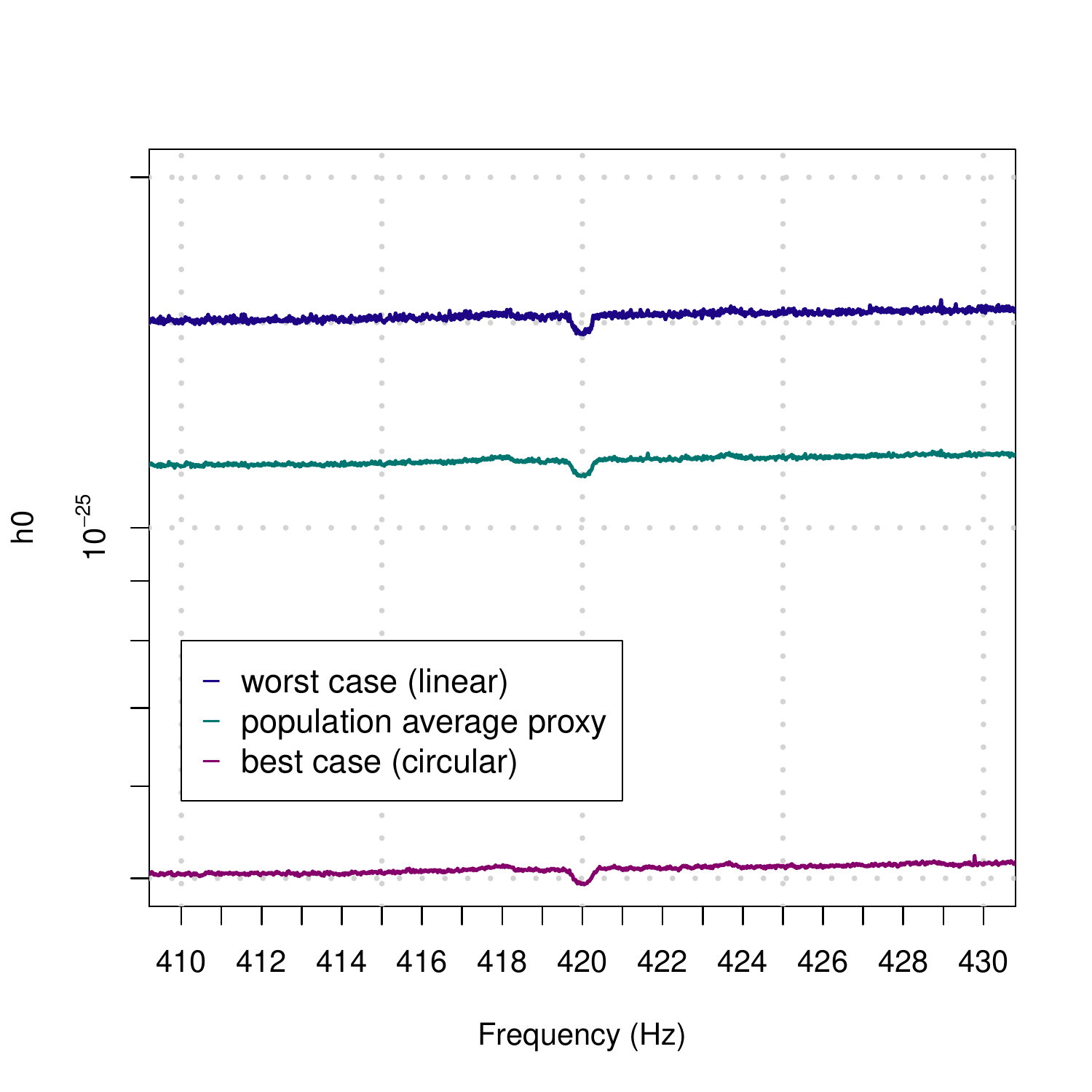}
\caption[Upper limits anomaly]{
\label{fig:anomaly}
One of anomalies in our upper limits due to errors in public data. At these high frequencies the shot noise is dominant. A 60 Hz power line harmonic can add to it, but a dip {\em below} the shot noise is completely unphysical. We suspect that the origin of this type of feature is the cleaning procedure applied to the publicly released data. We would very much like the opportunity to understand and fully characterize this effect, as we explain in the text. 
}
\end{figure}

\section{Anomalies in O2 data} 
 
The O2 LIGO data have been cleaned. The procedure removes a number of instrumental artifacts by regressing against instrumental channels. The original implementation \cite{O2_cleaning} was fairly conservative and was designed for searches of transient signals. 

The procedure was then applied more aggressively using a significantly higher number of instrumental channels. 
In this type of procedure overfitting is a danger. One of the signs of overfitting is the appearance of artificially low values in the cleaned data. We see examples of this in \cite{lvc_O2_allsky} in the upper limit and in the power spectral density plots. The presence of these unphysical data points is known \cite{o2_linespecs}. We find such systematic anomalies also in the upper limits from this search, Figure~\ref{fig:anomaly}. An argument that the low data values at certain frequencies are unphysical can also be made based on general grounds: at high frequencies the main contribution to the noise floor is shot noise, and regressing against auxiliary channels recording measurements from, say,  magnetometers and microphones, cannot reduce the noise below this level. 

Our pipeline establishes strict upper limits, within the stated calibration uncertainty \cite{Cahillane:2017vkb}. If ``overcleaning" happens, it will also affect a signal in the data, and compromise our upper limit results, unless the calibration uncertainty is correspondingly increased. Actually, all upper-limit methods need to know what any systematic from the cleaning procedure is, over the searched frequency range. 

A systematic underestimation of the noise level is also problematic when considering the search algorithms. Gravitational wave searches usually combine separate measurements with weights that down-weigh noisier periods. Data with artificially low noise skew the weights and give more importance to unphysical data. The effect shown in Figure~\ref{fig:anomaly} is small and, if that is all that happens, we could simply excise affected bands. 

In general we are concerned that whatever is causing the low-noise anomalies, has not been fully characterized and understood and hence could be affecting the data in subtle ways, without leaving such a visible signature, but still impacting the search results. If the auxiliary channels used to produce the cleaned data were publicly available, we would try and characterize features and magnitude of this effect and fold it in our strict upper limit values. In the absence of this, we present our upper limits taking the data at face value, but with the caveats explained in this Section.

\section{Results}

\begin{table*}[htbp]
\begin{center}
\begin{tabular}{D{.}{.}{3}D{.}{.}{5}D{.}{.}{4}D{.}{.}{4}D{.}{.}{4}l}\hline
\multicolumn{1}{c}{SNR}   &  \multicolumn{1}{c}{$f$} & \multicolumn{1}{c}{$\dot{f}$} &  \multicolumn{1}{c}{$\RAJ$}  & \multicolumn{1}{c}{$\DECJ$} & Comment \\
\multicolumn{1}{c}{}	&  \multicolumn{1}{c}{Hz}	&  \multicolumn{1}{c}{pHz/s} & \multicolumn{1}{c}{degrees} & \multicolumn{1}{c}{degrees} & \\
\hline \hline
7417.8 & 265.57535 & -1.0 & 71.591 & -56.223 & Injection ip0\\
6383.7 & 31.42473 & -0.5 & 285.078 & -58.269 & Injection ip11\\
3712.5 & 52.80832 & -0.2 & 302.538 & -83.843 & Injection ip5\\
1895.0 & 26.33722 & -0.9 & 234.716 & 41.244 & Injection ip10\\
1497.4 & 108.85716 & 0.0 & 178.369 & -33.439 & Injection ip3\\
27.1 & 83.43998 & 0.3 & 233.838 & -41.127 & \\
27.0 & 59.99418 & 0.6 & 211.421 & -21.567 & 60 Hz line\\
25.2 & 60.02252 & -0.2 & 25.723 & 37.383 & 60 Hz line\\
25.1 & 80.40734 & -0.9 & 225.641 & 77.846 & Line in H1\\
21.9 & 465.58924 & 0.6 & 87.318 & 50.980 & \\
21.9 & 197.87552 & -0.8 & 276.984 & -62.431 & \\
21.6 & 400.01133 & 0.0 & 66.157 & 89.153 & Coincident lines\\
21.6 & 31.84219 & 0.6 & 272.835 & 67.366 & \\
21.5 & 449.50129 & 0.7 & 308.074 & -32.371 & \\
21.3 & 55.60198 & 0.7 & 218.116 & -68.806 & \\
21.2 & 160.49660 & 0.5 & 90.941 & 32.384 & Line in L1\\
21.0 & 140.26640 & 0.7 & 179.620 & -19.718 & Line in H1\\
21.0 & 357.19349 & -0.6 & 202.263 & -27.984 & \\
21.0 & 153.87728 & -0.2 & 29.374 & -53.913 & Line in H1\\
21.0 & 440.03835 & 0.5 & 301.795 & 23.622 & \\
20.8 & 31.70025 & -1.1 & 101.926 & -66.112 & Line in H1\\
20.8 & 260.10055 & -0.5 & 85.537 & 17.565 & \\
20.7 & 180.10461 & -0.8 & 122.697 & -77.297 & 60 Hz harmonic\\
20.6 & 307.28973 & 1.1 & 330.582 & 28.601 & Near L1 line\\
20.6 & 441.11365 & -0.1 & 343.747 & -58.935 & \\
20.6 & 442.38421 & -0.0 & 20.788 & -50.078 & \\
20.5 & 312.31246 & 0.6 & 284.115 & -42.329 & \\
20.5 & 427.73851 & 0.8 & 259.995 & -44.007 & \\
20.5 & 456.66685 & 0.8 & 161.618 & 67.993 & Broad line in H1\\
20.4 & 457.11691 & -0.0 & 147.143 & 44.056 & Broad line in H1\\
20.4 & 245.06138 & 1.0 & 49.632 & 12.941 & \\
20.3 & 66.89584 & 0.8 & 177.962 & -21.334 & \\
20.2 & 448.02463 & 0.8 & 263.539 & 46.220 & \\
20.2 & 153.97881 & -0.5 & 131.683 & -58.036 & Line in H1\\
20.2 & 459.31715 & 0.0 & 341.763 & 20.999 & \\
20.2 & 192.60299 & -0.6 & 340.182 & 50.884 & \\
20.1 & 394.38609 & -0.1 & 257.670 & -82.979 & \\
20.1 & 478.98702 & 0.3 & 151.593 & -67.970 & Line in L1\\
20.1 & 317.99684 & -1.0 & 26.269 & -22.945 & \\
20.0 & 436.12088 & -1.3 & 342.703 & -40.723 & \\

\hline
\end{tabular}
\caption[Outliers produced by the detection pipeline]{High SNR outliers produced by the detection pipeline. We list outliers with SNR above 20, and exclude those near ecliptic poles and associated with stationary lines. Only the highest-SNR outlier is shown for each 0.1\,Hz frequency region. Outliers marked ``ipX'' are due to simulated signals ``hardware-injected'' during the science run for validation purposes. Their parameters are listed in Table \ref{tab:injections}. Outliers marked with ``line'' have strong narrowband disturbances near the outlier frequency.
Signal frequencies refer to GPS epoch $1176425033$. Full list of outliers down to SNR$>16$ is available in \cite{data}.} 
\label{tab:Outliers}
\end{center}
\end{table*}

\begin{table}[htbp]
\begin{center}
\begin{tabular}{lD{.}{.}{6}cD{.}{.}{5}D{.}{.}{4}}
\hline
Label & \multicolumn{1}{c}{Frequency} & \multicolumn{1}{c}{Spindown} & \multicolumn{1}{c}{$\RAJ$} & \multicolumn{1}{c}{$\DECJ$} \\
 & \multicolumn{1}{c}{Hz} & \multicolumn{1}{c}{pHz/s} & \multicolumn{1}{c}{degrees} & \multicolumn{1}{c}{degrees} \\
\hline \hline

ip0   &  265.575343  & $-4.15$   &   71.55193     &  -56.21749 \\
ip3   &  108.857159  & $\sci{-1.46}{-5}$   &  178.37257     &  -33.4366  \\
ip5   & 52.808324  & $\sci{-4.03}{-6}$     &  302.62664     &  -83.83914 \\
ip6   &  145.860492 & $-6730$     &  358.75095     &  -65.42262 \\
ip8   &  190.634274 & $-8650$     &  351.38958     &  -33.41852 \\
ip10  &   26.338016 & $-85$    &  221.55565     &   42.87730 \\
ip11  &   31.424735 & $-0.507$    &  285.09733     &  -58.27209 \\
ip12  &   38.764787 & $-6250$     &  331.85267     &  -16.97288 \\
\hline
\end{tabular}
\caption[Parameters of hardware injections]{Parameters of the hardware-injected simulated continuous wave signals during the O2 data run (GPS epoch $1176425033$).}
\label{tab:injections}
\end{center}
\end{table}

Based on the search results, we constrain the intrinsic strength $h_0$ of signals at the detectors. The $h_0$ upper limits also hold if a signal is present at a detectable level. In fact, in the band hosting loud outliers, for example due to the hardware injections, the upper limit values are higher than elsewhere.

In order to present the results in a concise way, each upper limit holds for signals with frequency from a certain frequency band, the whole sky and with spin-down values within the search range. This is completely standard practice in presenting continuous wave search results \cite{EHO2,lvc_O2_allsky,O2_falcon2}. Details on the specific procedure that we use -- the universal statistics algorithm -- may be found in \cite{universal_statistics}; here we recap the main features. 

The universal statistics algorithm establishes robust upper limits on the power of a single putative signal given a set of independent power measurements. The sets of power measurements are computed after filtering the data for each combination of frequency evolution parameters, sky position and polarization.  Each measurement is then associated with 95\% confidence level upper limit on signal amplitude. Because this procedure does not require injections it is computationally efficient to iterate over the entire parameter space. However, the volume of data produced is very large and the upper limits are aggregated by taking the maximum over marginalized parameters. A correction is applied to account for mismatch due to parameter space sampling. The upper limits are verified with Monte-Carlo injections.

The upper limit procedure is performed in narrow bands assuming that the input data have been perfectly calibrated. In reality, the calibration of the input data can be wrong in amplitude and phase \cite{Cahillane:2017vkb}, but this varies slowly over frequency and can be assumed constant in each upper-limit band. The amplitude uncertainty simply translates in an uncertainty in the upper limit of the same magnitude, while the phase uncertainty is negligible.

Because we take the least favourable sky position and polarisation, the upper limit is the ``worst-case" upper limit, and generally a signal will not need to be that loud to match our observation. The worst-case upper limits are in this sense conservative, but it is precisely because of this that they are strict upper limits, that hold true for any signal parameter combination. We also quote a ``best-case" upper limit that corresponds to a population of circularly polarized signals, but still from the least favourable sky position. 

It is common in the literature \cite{EHO2,lvc_O2_allsky} to compute instead population-average upper limits, which do not hold at the given confidence for the least favourable signal parameter combinations. In order to ease comparison of our results with other methods, we compute a population-average proxy using a weighted sum of polarization-specific upper limits. We found that signals injected at the population average proxy level have between 90-95\% chance of being recovered depending on the contamination of the underlying data by detector artifacts.

The gravitational wave amplitude upper limits from this search are shown in Figure \ref{fig:amplitudeULs} and are available in machine readable format at \cite{data}. 

Our upper limits can be translated into limits on gravitational waves from  boson condensates around black holes \cite{boson1,boson2}, which are expected to emit monochromatic continuous wave signals \cite{discovering_axions} with very small frequency drift. We leave it to the interested reader to constrain from our upper limits physical quantities of interest, based on the specific model they wish to consider (for example, an ensemble signal of \cite{Zhu:2020tht}).

The search also identifies small areas of parameter space containing waveforms that are unusually consistent with the data. These are the outliers of the search. The outliers are often caused by signals of terrestrial origin due to technical or environmental noise sources; the weaker ones may also be fluctuations. A key result of our search is a number of interesting outliers (Table~\ref{tab:Outliers}), many of which are located in bands with clean frequency spectrum. Table~\ref{tab:Outliers} only lists outliers with SNR above 20 and excludes outliers near ecliptic pole most likely associated with instrumental lines, in particular the outliers from the 0.5\,Hz frequency comb. The full list of outliers down to SNR~16 is available in \cite{data}.

Monte Carlo simulations show that we detect IT2 signals within $\sci{8}{-6}$\,Hz of the signal frequency $f$, within $10^{-12}$\,Hz/s of its frequency derivative, and within $0.12\textrm{\,Hz}/f$ radians from its sky location, the last calculated after projection on the ecliptic plane (``ecliptic distance''). For non-IT2 signals, or signals with frequency derivative outside search range, the tolerances should be widened. In particular, there is a degeneracy between sky location and frequency derivative, so a mismatch in the latter results in outlier offset from its true position on the sky.

Because the coherence length of our followup stage has been extended to 16\,days, we have much tighter tolerance for the signals that pass our pipeline compared with previous searches. While we see 5 simulated signals clearly (ip0, ip3, ip5, ip10 and ip11, Table \ref{tab:injections}), the injections ip6 and ip12 do not produce any outliers as their frequency derivative smaller than $\sci{-5}{-9}$\,Hz/s is well beyond our target parameter space. ip8 is loud enough to induce weak outliers with SNR below 20 even though its spindown is also outside of our target parameter space. The ip8 outliers are included in the supplemental materials file.

\section{Conclusions}

We have performed the most sensitive search for continuous gravitational waves in the frequency band of 20-500\,Hz. Just as in our previous papers \cite{O2_falcon, O2_falcon2}, we have a number of interesting outliers. Our constraints are tighter than before, the signal-to-noise ratio is higher. Are they due to some sort of noise? Or are we seeing astrophysical sources? The latter is unlikely if we assume a classic IT2 source with ellipticity of $10^{-8}$ - we simply do not expect that many sources within 44\,pc. Even widening the reach to 440\,pc by including ellipticities of $10^{-7}$, it would be surprising that this many outliers are of astrophysical origin. 

In order to make solid prediction on the number of detectable sources, one would need to have a mechanism that guarantees equatorial ellipticity in some minimal number of neutron stars rotating between a few tens and a few hundreds of Hz. We are not aware of any papers that provide such lower bounds, though there are efforts to characterise source populations under various assumptions \cite{population1,population2}. On the other hand, based on the null results of many different continuous wave searches, including all-sky surveys, directed searches and targeted searches - the latter probing significantly lower gravitational wave amplitude values than an all-sky search -  a reasonable aspiration for results from an all-sky search is to find one signal.

We look forward to more data for further analysis.

\begin{acknowledgments}
The authors thank the scientists, engineers and technicians of LIGO, whose hard work and dedication produced the data that made this search possible.

The search was performed on the ATLAS cluster at AEI Hannover. We thank Bruce Allen, Carsten Aulbert and Henning Fehrmann for their support.

This research has made use of data, software and/or web tools obtained from the LIGO Open Science Center (\url{https://losc.ligo.org}), a service of LIGO Laboratory, the LIGO Scientific Collaboration and the Virgo Collaboration.  LIGO is funded by the U.S. National Science Foundation. Virgo is funded by the French Centre National de Recherche Scientifique (CNRS), the Italian Istituto Nazionale della Fisica Nucleare (INFN) and the Dutch Nikhef, with contributions by Polish and Hungarian institutes.
\end{acknowledgments}

\end{document}